\documentclass[11pt]{article}%
\usepackage{amsfonts}
\usepackage{amsmath}
\usepackage{amssymb}
\usepackage{graphicx}%
\providecommand{\U}[1]{\protect\rule{.1in}{.1in}}
\newtheorem{theorem}{Theorem}

\newtheorem{notation}[theorem]{Notation}

\begin{document}

\title{FERMI'S TRICK\ AND\ SYMPLECTIC\ CAPACITIES: A\ GEOMETRIC\ PICTURE\ OF\ QUANTUM\ STATES}
\author{Maurice A. de Gosson\thanks{This work has been supported by the Austrian
Research Agency FWF (Projektnummer P20442-N13).}\\\ \textit{University of Vienna }\\\textit{Faculty of Mathematics, NuHAG}\\\textit{A-1090 Vienna, AUSTRIA}
\and Serge M. de Gosson\\\textit{Swedish Social Insurance Agency}\\\textit{Department for Analysis and Forecasts}\\\textit{103 51 Stockholm}}
\maketitle

\begin{abstract}
We extend the notion of \textquotedblleft quantum blob\textquotedblright%
\ studied in previous work to excited states of the generalized harmonic
oscillator in $n$ dimensions. This extension is made possible by Fermi's
observation in 1930 that the state of a quantum system may be defined in two
different (but equivalent) ways, namely by its wavefunction $\Psi$ or by a
certain function $g_{\mathrm{F}}$ on phase space canonically associated with
$\Psi$. We study Fermi's function when $\Psi$ is a Gaussian (generalized
coherent state). A striking result is that we can use the Ekeland--Hofer
symplectic capacities to characterize the Fermi functions of the excited
states of the generalized harmonic oscillator, leading to new insight on the
relationship between symplectic topology and quantum mechanics.

\end{abstract}

\section{Introduction\label{secfe}}

\subsection{What We Want to Do}

We address the question whether it is possible to represent geometrically a
function $\psi$ of the variables $x=(x_{1},x_{2},..,x_{n})$. The problem is in
fact easy to answer if $\psi$ is a Gaussian function because then its Wigner
transform is proportional to a Gaussian $e^{-\frac{1}{\hbar}S^{T}Sz\cdot z}$
where $S$ is a symplectic matrix uniquely determined by $\psi$. It follows
that there is a one-to-one correspondence between Gaussians and the sets
$S^{T}Sz\cdot z\leq\hbar$. We have called these sets \textquotedblleft quantum
blobs\textquotedblright\ in \cite{de02-2,de03-2,de04,de05,Birk,go09,degostat};
the interest of these quantum blobs comes from the fact that they represent
minimum uncertainty sets in phase space.

The Gaussian function
\begin{equation}
\Psi_{0}(x)=e^{-x^{2}/2\hbar}; \label{fid}%
\end{equation}
is the (unnormalized) ground state of the one-dimensional harmonic oscillator
with mass and frequency equal to one: $\widehat{H}\Psi_{0}=E_{0}\Psi_{0}$
where $E_{0}=\tfrac{1}{2}\hbar$ and
\begin{equation}
\widehat{H}=\frac{1}{2}\left(  -\hbar^{2}\frac{d^{2}}{dx^{2}}+x^{2}\right)
\label{gf5}%
\end{equation}
This operator is the quantization of the classical oscillator Hamiltonian%
\begin{equation}
H(x,p)=\frac{1}{2}(p^{2}+x^{2}) \label{gf7}%
\end{equation}
The set $\Omega_{0}$ defined by the inequality $H\leq E_{0}$ is the interior
of the energy hypersurface $H\leq E_{0}$; it is the disk $p^{2}+x^{2}\leq
\hbar$ with radius $R_{0}=\sqrt{\hbar}$. Let us now consider the $N$-th
excited state of the operator $\widehat{H}$; it is the (unnormalized) Hermite
function%
\begin{equation}
\Psi_{N}(x)=e^{-x^{2}/2\hbar}H_{N}(x/\sqrt{\hbar}) \label{hermite1}%
\end{equation}
where
\begin{equation}
H_{N}(x)=(-1)^{n}e^{x^{2}}\tfrac{d^{N}}{dx^{N}}e^{-x^{2}} \label{hermite}%
\end{equation}
is the $N$-th Hermite polynomial. It is a solution of $\widehat{H}\Psi
_{N}=\left(  N+\tfrac{1}{2}\right)  \hbar\Psi_{N}$ and the set $\Omega_{N}$
defined by the inequality $H\leq E_{N}=(2N+1)\hbar$ is again a disk, but this
time with radius $R_{N}=\sqrt{\left(  N+\frac{1}{2}\right)  h}$.

In this paper we introduce a non-trivial extension of the notion of
\textquotedblleft quantum blob\textquotedblright\ we defined and studied in
previous work. Quantum blobs are deformations of the phase space ball
$|x|^{2}+|p|^{2}\leq\hbar$ by translations and linear canonical
transformations. Their interest come from the fact that they provide us with a
coarse-graining of phase space different from the usual coarse graining by
cubes with volume $\sim h^{n}$ commonly used in statistical mechanics. They
appear as space units of minimum uncertainty in one-to-one correspondence with
the generalized coherent states familiar from quantum optics, and have allowed
us to recover the exact ground states of generalized harmonic oscillators, as
well as the semiclassical energy levels of quantum systems with completely
integrable Hamiltonian function, and to explain them in terms of the
topological notion of symplectic capacity \cite{HZ,Polterovich} originating in
Gromov's \cite{Gromov} non-squeezing theorem (alias \textquotedblleft the
principle of the symplectic camel\textquotedblright). Quantum blobs, do not,
however, allow a characterization of excited states; for instance there is no
obvious relation between them and the Hermite functions. Why this does not
work is easy to understand: quantum blobs correspond to the states saturating
the Schr\"{o}dinger--Robertson inequalities%
\begin{equation}
(\Delta X_{j})^{2}(\Delta P_{j})^{2}\geq\Delta(X_{j},P_{j})^{2}+\tfrac{1}%
{4}\hbar^{2}\text{ , }1\leq j\leq n;\label{RS}%
\end{equation}
as is well-known \cite{degosat} the quantum states for which all these
inequalities become equalities are Gaussians, in this case precisely those who
are themselves the ground states of generalized harmonic oscillators. As soon
as one consider the excited states the corresponding eigenfunctions are
Hermite functions and for these the inequalities (\ref{RS}) are strict. The
way out of this difficulty is to define new phase space objects, the
\textquotedblleft Fermi blobs\textquotedblright\ of the title of this paper.
Such an approach should certainly be welcome in times where phase space is
beginning to be taken seriously (see the recent review paper \cite{new}).

\subsection{How We Will Do It}

We will show that a complete geometric picture of excited states can be given
using an idea of the physicist Enrico Fermi in a largely forgotten paper
\cite{Fermi} from 1930. Fermi associates to every quantum state $\Psi$ a
certain hypersurface $g_{\mathrm{F}}(x,p)=0$ in phase space. The underlying
idea is actually surprisingly simple. It consists in observing that any
complex twice continuously differentiable function $\Psi(x)=R(x)e^{i\Phi
(x)/\hslash}$ ($R(x)\geq0$ and $\Phi(x)$ real) defined on $\mathbb{R}^{n}$
satisfies the partial differential equation%
\begin{equation}
\left[  \left(  -i\hbar\nabla_{x}-\nabla_{x}\Phi\right)  ^{2}+\hbar^{2}%
\frac{\nabla_{x}^{2}R}{R}\right]  \Psi=0.\label{gf1}%
\end{equation}
where $\nabla_{x}^{2}$ is the Laplace operator in the variables $x_{1}%
,...,x_{n}$ (it is assumed that $R(x)\neq0$ for $x$ in some subset of
$\mathbb{R}^{n}$). Performing the gauge transformation $-i\hbar\nabla
_{x}\longrightarrow-i\hbar\nabla_{x}-\nabla_{x}\Phi$, this equation is in fact
equivalent to the trivial equation%
\begin{equation}
\left(  -\hbar^{2}\nabla_{x}^{2}+\hbar^{2}\frac{\nabla_{x}^{2}R}{R}\right)
R=0.\label{trivial}%
\end{equation}
The operator
\begin{equation}
\widehat{g_{\mathrm{F}}}=\left(  -i\hbar\nabla_{x}-\nabla_{x}\Phi\right)
^{2}+\hbar^{2}\frac{\nabla_{x}^{2}R}{R}\label{fermop}%
\end{equation}
appearing in the left-hand side of Eqn. (\ref{gf1}) is the quantisation (in
every reasonable physical quantisation scheme) of the real observable
\begin{equation}
g_{\mathrm{F}}(x,p)=\left(  p-\nabla_{x}\Phi\right)  ^{2}+\hbar^{2}%
\frac{\nabla_{x}^{2}R}{R}\label{gf2}%
\end{equation}
and the equation $g_{\mathrm{F}}(x,p)=0$ in general determines a hypersurface
$\mathcal{H}_{\mathrm{F}}$ in phase space $\mathbb{R}_{x,p}^{2n}$ which Fermi
ultimately \emph{identifies} with the state $\Psi$ itself. The remarkable
thing with this construction is that it shows that to an arbitrary function
$\Psi$ it associates a Hamiltonian function of the classical type%
\begin{equation}
H=\left(  p-\nabla_{x}\Phi\right)  ^{2}+V\label{classical}%
\end{equation}
even if $\Psi$ is the solution of another partial (or pseudo-differential)
equation. We notice that when $\Psi$ is an eigenstate of the operator
$\widehat{H}\Psi=E\Psi$ then $g_{\mathrm{F}}=H-E$ and $\mathcal{H}%
_{\mathrm{F}}$ is just the energy hypersurface $H(x,p)=E$.

Of course, Fermi's analysis was very heuristic and its mathematical rigour
borders the unacceptable (at least by modern standards). Fermi's paper has
recently been rediscovered by Benenti \cite{benenti} and Benenti and Strini
\cite{best}, who study its relationship with the level sets of the Wigner
transform of $\Psi$.

\begin{notation}
The points in configuration and momentum space are written $x=(x_{1}%
,...,x_{n})$ and $p=(p_{1},...,p_{n})$ respectively; in formulas $x$ an $p$
are viewed as column vectors. We will also use the collective notation
$z=(x,p)$ for the phase space variable. The matrix $J=%
\begin{pmatrix}
0 & I\\
-I & 0
\end{pmatrix}
$ ($0$ and $I$ the $n\times n$ zero and identity matrices) defines the
standard symplectic form on the phase space $\mathbb{R}_{x}^{2n}$ via the
formula $\sigma(z,z^{\prime})=Jz\cdot z^{\prime}=p\cdot x^{\prime}-p^{\prime
}\cdot x$. We write $\hbar=h/2\pi$, $h$ being Planck's constant. The
symplectic group is denoted by $\operatorname*{Sp}(2n,\mathbb{R})$: it is the
multiplicative group of all real $2n\times2n$ matrices $S$ such that
$\sigma(Sz,Sz^{\prime})=\sigma(z,z^{\prime})$ for all $z,z^{\prime}$.
\end{notation}

\section{Symplectic Capacities and Quantum Blobs}

To generalize the discussion above to the multi-dimensional case we have to
introduce some concepts from symplectic topology. For a review of these
notions\ see de Gosson and Luef \cite{golu10}.

\subsection{Symplectic Capacities}

\subsubsection*{Intrinsic symplectic capacities}

An \textit{intrinsic} symplectic capacity assigns a non-negative number (or
$+\infty$) $c(\Omega)$ to every subset $\Omega$ of phase space $\mathbb{R}%
^{2n}$; this assignment is subjected to the following properties:

\begin{itemize}
\item \textbf{Monotonicity:} If $\Omega\subset\Omega^{\prime}$ then
$c(\Omega)\leq c(\Omega^{\prime})$;

\item \textbf{Symplectic invariance:} If $f$ is a canonical transformation
(linear, or not) then $c(f(\Omega))=c(\Omega)$;

\item \textbf{Conformality:} If $\lambda$ is a real number then $c(\lambda
\Omega)=\lambda^{2}c(\Omega)$; here $\lambda\Omega$ is the set of all points
$\lambda z$ when $z\in\Omega$;

\item \textbf{Normalization:} We have
\begin{equation}
c(B^{2n}(R))=\pi R^{2}=c(Z_{j}^{2n}(R)); \label{norm1}%
\end{equation}
here $B^{2n}(R)$ is the phase-space ball $|x|^{2}+|p|^{2}\leq R^{2}$ and
$Z_{j}^{2n}(R)$ the phase-space cylinder $x_{j}^{2}+p_{j}^{2}\leq R^{2}$.
\end{itemize}

Let $c$ be a symplectic capacity on the phase plane $\mathbb{R}^{2}$. We have
$c(\Omega)=\operatorname*{Area}(\Omega)$ when $\Omega$ is a connected and
simply connected surface. In the general case there exist infinitely many
intrinsic symplectic capacities, but they all agree on phase space ellipsoids
as we will see below. The smallest symplectic capacity is denoted by $c_{\min
}$ (\textquotedblleft Gromov width\textquotedblright): by definition $c_{\min
}(\Omega)$ is the supremum of all numbers $\pi R^{2}$ such that there exists a
canonical transformation such that $f(B^{2n}(R))\subset\Omega$. The fact that
$c_{\min}$ really is a symplectic capacity follows from a deep and difficult
topological result, Gromov's \cite{Gromov} symplectic non-squeezing theorem,
alias the principle of the symplectic camel. (For a discussion of Gromov's
theorem from the point of view of Physics see de Gosson \cite{go09}, de Gosson
and Luef \cite{golu10}.) Another useful example is provided by the
Hofer--Zehnder \cite{HZ} capacity $c^{\mathrm{HZ}}$. It has the property that
it is given by the integral of the action form $pdx=p_{1}dx_{1}+\cdot
\cdot\cdot+p_{n}dx_{n}$ along a certain curve:%
\begin{equation}
c^{\text{HZ}}(\Omega)=\oint\nolimits_{\gamma_{\min}}pdx \label{chz}%
\end{equation}
when $\Omega$ is a compact convex set in phase space; here $\gamma_{\min}$ is
the shortest (positively oriented) Hamiltonian periodic orbit carried by the
boundary $\partial\Omega$ of $\Omega$. This formula agrees with the usual
notion of area in the case $n=1$.

It turns out that all intrinsic symplectic capacities agree on phase space
ellipsoids, and are calculated as follows (see e.g. \cite{Birk,golu10,HZ}).
Let $M$ be a $2n\times2n$ positive-definite matrix $M$ and consider the
ellipsoid:%
\begin{equation}
\Omega_{M,z_{0}}:M(z-z_{0})^{2}\leq1. \label{ellipsoid}%
\end{equation}

Then, for every intrinsic symplectic capacity $c$ we have
\begin{equation}
c(\Omega_{M,z_{0}})=\pi/\lambda_{\max}^{\sigma} \label{capellipse}%
\end{equation}
where $\lambda_{\max}^{\sigma}=$ is the largest symplectic eigenvalue of $M$.
The symplectic eigenvalues of a positive definite matrix are defined as
follows: the matrix $JM$ ($J$ the standard symplectic matrix) is equivalent to
the antisymmetric matrix $M^{1/2}JM^{1/2}$ hence its $2n$ eigenvalues are of
the type $\pm i\lambda_{1}^{\sigma},..,$ $\pm i\lambda_{n}^{\sigma}$ where
$\lambda_{j}^{\sigma}>0$. The positive numbers $\lambda_{1}^{\sigma},..,$
$\lambda_{n}^{\sigma}$ are called the \emph{symplectic eigenvalues} of the
matrix $M$.

In particular, if $X$ and $Y$ are real symmetric $n\times n$ matrices, then
the symplectic capacity of the ellipsoid%
\begin{equation}
\Omega_{(A,B)}:Xx^{2}+Yp^{2}\leq1 \label{capab}%
\end{equation}
is given by%
\begin{equation}
c(\Omega_{(A,B)})=\pi/\sqrt{\lambda_{\max}} \label{cab}%
\end{equation}
where $\lambda_{\max}$ is the largest eigenvalue of $AB$.

\subsubsection*{Extrinsic symplectic capacities}

The definition of an extrinsic symplectic capacity is similar to that of an
intrinsic capacity, but one weakens the normalization condition (\ref{norm1})
by only requiring that:

\begin{itemize}
\item \textbf{Nontriviality:} $c(B^{2n}(R))<+\infty$ and $c(Z_{j}%
^{2n}(R))<+\infty$.
\end{itemize}

In \cite{EH} Ekeland and Hofer defined a sequence $c_{1}^{\mathrm{EH}}$,
$c_{2}^{\mathrm{EH}},...,c_{k}^{\mathrm{EH}},...$ of extrinsic symplectic
capacities satisfying the nontriviality properties%
\begin{equation}
c_{k}^{\mathrm{EH}}(B^{2n}(R))=\left[  \frac{k+n-1}{n}\right]  \pi
R^{2}\ \ \text{,}\ \ c_{k}^{\mathrm{EH}}(Z_{j}^{2n}(R))=k\pi R^{2}.\label{eh}%
\end{equation}
Of course $c_{1}^{\mathrm{EH}}$ is an intrinsic capacity; in fact it coincides
with the Hofer--Zehnder capacity on bounded convex sets $\Omega$. We have%
\begin{equation}
c_{1}^{\text{EH}}(\Omega)\leq c_{2}^{\text{EH}}(\Omega)\leq\cdot\cdot\cdot\leq
c_{k}^{\text{EH}}(\Omega)\leq\cdot\cdot\cdot
\end{equation}
The Ekeland--Hofer capacities have the property that for each $k$ there exists
an integer $N\geq0$ and a closed characteristic $\gamma$ of $\partial\Omega$
such that%
\begin{equation}
c_{k}^{\text{EH}}(\Omega)=N\left\vert \oint\nolimits_{\gamma}pdx\right\vert
\label{acspec}%
\end{equation}
(in other words, $c_{k}^{\text{EH}}(\Omega)$ is a value of the \textit{action
spectrum} \cite{quasyge} of the boundary $\partial\Omega$ of $\Omega$); this
formula shows that $c_{k}^{\text{EH}}(\Omega)$ is solely determined by
$\partial\Omega$; therefore the notation $c_{k}^{\text{EH}}(\partial\Omega)$
is often used in the literature. The Ekeland--Hofer capacities $c_{k}%
^{\text{EH}}$ allow us to classify phase-space ellipsoids. In fact, the
non-decreasing sequence of numbers $c_{k}^{\text{EH}}(\Omega_{M})$ is
determined as follows for an ellipsoid $\Omega:Mz\cdot z\leq1$ ($M$ symmetric
and positive-definite): let \ $(\lambda_{1}^{\sigma},...,\lambda_{n}^{\sigma
})$ be the symplectic eigenvalues of $M$; then
\begin{equation}
\{c_{k}^{\text{EH}}(\Omega):k=1,2,...\}=\{N\pi\lambda_{j}^{\sigma
}:j=1,...,n;N=0,1,2,...\}.\label{cehc}%
\end{equation}
Equivalently, the increasing sequence $c_{1}^{\text{EH}}(\Omega)\leq
c_{2}^{\text{EH}}(\Omega)\leq\cdot\cdot\cdot$ is obtained by writing the
numbers $N\pi\lambda_{j}^{\sigma}$ in increasing order with repetitions if a
number occurs more than once.

\subsection{Quantum Blobs}

By definition a quantum blob $\mathcal{QB}^{2n}(z_{0},S)$ is the image of the
phase space ball $B^{2n}(S^{-1}z_{0},\sqrt{\hbar}):|z-S^{-1}z_{0}|\leq
\sqrt{\hbar}$ by a linear canonical transformation (identified with a
symplectic matrix $S$). A quantum blob is thus a phase space ellipsoid with
symplectic capacity $\pi\hbar=\frac{1}{2}h$, but it is not true that,
conversely, an arbitrary phase space ellipsoid with symplectic capacity
$\frac{1}{2}h$ is a quantum blob. One can however show (de Gosson
\cite{de04,de05,Birk}, de Gosson and Luef \cite{golu10}) that such an
ellipsoid contains a unique quantum blob. One proves (ibid.) that a quantum
blob $\mathcal{QB}^{2n}(z_{0},S)$ is characterized by the two following
\emph{equivalent} properties:

\begin{itemize}
\item \textit{The intersection of the ellipsoid }$\mathcal{QB}^{2n}(z_{0}%
,S)$\textit{\ with a plane passing through }$z_{0}$\textit{\ and parallel to
any of the plane of canonically conjugate coordinates }$x_{j},p_{j}%
$\textit{\ in }$\mathbb{R}_{z}^{2n}$ \textit{is an ellipse with area }%
$\frac{1}{2}h$\textit{; }

\item \textit{The supremum of the set of all numbers }$\pi R^{2}%
$\textit{\ such that the ball }$B^{2n}(\sqrt{R}):|z|\leq R$\textit{\ can be
embedded into }$\mathcal{QB}^{2n}(z_{0},S)$\textit{\ using canonical
transformations (linear, or not) is }$\frac{1}{2}h$\textit{. Hence no phase
space ball with radius }$R>\sqrt{\hbar}$ \textit{can be \textquotedblleft
squeezed\textquotedblright\ inside }$\mathcal{QB}^{2n}(z_{0},S)$%
\textit{\ using only canonical transformations.}
\end{itemize}

It turns out (de Gosson \cite{Birk}) that in the first of these conditions one
can replace the plane of conjugate coordinates with any symplectic plane (a
symplectic plane is a two-dimensional subspace of $\mathbb{R}_{z}^{2n}$ on
which the restriction of the symplectic form $\sigma$ is again a symplectic
form). There is a natural action%
\[
\operatorname*{Sp}(2n,\mathbb{R})\times\mathcal{QB}(2n,\mathbb{R}%
)\longrightarrow\mathcal{QB}(2n,\mathbb{R})
\]
of the symplectic group on quantum blobs.

\section{Generalized Coherent States}

\subsection{The Fermi Function of a Gaussian}

We next consider arbitrary (normalized) generalized coherent states%
\begin{equation}
\Psi_{X,Y}(x)=\left(  \frac{1}{\pi\hbar}\right)  ^{n/4}(\det X)^{1/4}%
\exp\left[  -\frac{1}{2\hbar}(X+iY)x\cdot x\right]  \label{coh1}%
\end{equation}
where $X$ and $Y$ are real symmetric $n\times n$ matrices, and $X$ is positive
definite. Setting $\Phi(x)=-\frac{1}{2}Yx\cdot x$ and $R(x)=\exp\left(
-\frac{1}{2\hbar}Xx\cdot x\right)  $ we have
\begin{equation}
\nabla_{x}\Phi(x)=-Yx\text{ \ , \ }\frac{\nabla_{x}^{2}R(x)}{R(x)}=-\frac
{1}{\hbar}\operatorname*{Tr}X+\frac{1}{\hbar^{2}}X^{2}x\cdot x\label{tr}%
\end{equation}
hence the Fermi function of $\Psi_{X,Y}$ is the quadratic form
\begin{equation}
g_{\mathrm{F}}(x,p)=(p+Yx)^{2}+X^{2}x\cdot x-\hbar\operatorname*{Tr}%
X.\label{gf3}%
\end{equation}
We can rewrite this formula as
\begin{equation}
g_{\mathrm{F}}(x,p)=M_{\mathrm{F}}z\cdot z-\hbar\operatorname*{Tr}X\label{gf}%
\end{equation}
($z=(x,p)$) where $M_{\mathrm{F}}$ is the symmetric matrix
\begin{equation}
M_{\mathrm{F}}=%
\begin{pmatrix}
X^{2}+Y^{2} & Y\\
Y & I
\end{pmatrix}
.\label{mf}%
\end{equation}
A straightforward calculation shows that we have the factorization%
\begin{equation}
M_{\mathrm{F}}=S^{T}%
\begin{pmatrix}
X & 0\\
0 & X
\end{pmatrix}
S\label{mfs}%
\end{equation}
where $S$ is the \emph{symplectic} matrix
\begin{equation}
S=%
\begin{pmatrix}
X^{1/2} & 0\\
X^{-1/2}Y & X^{-1/2}%
\end{pmatrix}
.\label{ess}%
\end{equation}

It turns out --and this is really a striking fact!-- that $M_{\mathrm{F}}$ is
closely related to the Wigner transform
\begin{equation}
W\Psi_{X,Y}(z)=\left(  \frac{1}{2\pi\hbar}\right)  ^{n}\int_{\mathbb{R}^{n}%
}e^{-\frac{i}{\hbar}p\cdot y}\Psi_{X,Y}(x+\tfrac{1}{2}y)\Psi_{X,Y}^{\ast
}(x-\tfrac{1}{2}y)dy\label{oupsi}%
\end{equation}
of the state $\Psi_{X,Y}$ because we have%
\begin{equation}
W\Psi_{X,Y}(z)=\left(  \frac{1}{\pi\hbar}\right)  ^{n}\exp\left(  -\frac
{1}{\hbar}Gz\cdot z\right)  \label{goupsi}%
\end{equation}
where $G$ is the symplectic matrix%
\begin{equation}
G=S^{T}S=%
\begin{pmatrix}
X+YX^{-1}Y & YX^{-1}\\
X^{-1}Y & X^{-1}%
\end{pmatrix}
\label{G}%
\end{equation}
(see e.g. \cite{Birk,Littlejohn}). When $n=1$ and $\Psi_{X,Y}(x)=\Psi_{0}(x)$
the fiducial coherent state (\ref{fid}) we have $S^{-1}D^{-1/2}S=I$ and
$\operatorname*{Tr}X=1$ hence the formula%
\[
W\Psi_{0}(z)=\left(  \frac{1}{\pi\hbar}\right)  ^{1/4}\frac{1}{e}\exp\left[
-\frac{1}{\hbar}M_{\mathrm{F}}z\cdot z\right]
\]
already observed by Benenti and Strini in \cite{best}.

\subsection{\label{subsecgi}Geometric Interpretation}

Recall (formula (\ref{capellipse})) that the symplectic capacity $c(\Omega)$
of an ellipsoid $Mz\cdot z\leq1$ ($M$ a symmetric positive definite
$2n\times2n$ matrix) is given by
\begin{equation}
c(\Omega)=\pi/\lambda_{\max}^{\sigma}\label{cw}%
\end{equation}
where $\lambda_{\max}^{\sigma}=\max\{\lambda_{1}^{\sigma},..,$ $\lambda
_{n}^{\sigma}\}$, the $\lambda_{j}^{\sigma}$ being the symplectic eigenvalues
of $M$. We denote by $\Omega_{\mathrm{F}}$ the phase space ellipsoid defined
by $g_{\mathrm{F}}(x,p)\leq0$, that is:\
\[
\Omega_{\mathrm{F}}:M_{\mathrm{F}}z\cdot z\leq\hbar\operatorname*{Tr}X;
\]
it is the ellipsoid bounded by the Fermi hypersurface $\mathcal{H}%
_{\mathrm{F}}$ corresponding to the generalized coherent state $\Psi_{X,Y}$.
Let us perform the symplectic change of variables $z^{\prime}=Sz$; in the new
coordinates the ellipsoid $\Omega_{\mathrm{F}}$ is represented by the
inequality
\begin{equation}
Xx^{\prime}\cdot x^{\prime}+Xp^{\prime}\cdot p^{\prime}\leq\hbar
\operatorname*{Tr}X\label{ferx}%
\end{equation}
hence $c(\Omega_{\mathrm{F}})$ equals the symplectic capacity of the ellipsoid
(\ref{ferx}). Applying the rule above we thus have to find the symplectic
eigenvalues of the block-diagonal matrix $%
\begin{pmatrix}
X & 0\\
0 & X
\end{pmatrix}
$; a straightforward calculation shows that these are just the eigenvalues
$\omega_{1},...,\omega_{n}$ of $X$ and hence%
\begin{equation}
c(\Omega_{\mathrm{F}})=\pi\hbar\operatorname*{Tr}X/\omega_{\max}\label{cwf}%
\end{equation}
where $\omega_{\max}=\max\{\omega_{1},...,\omega_{n}\}$. In view of the
trivial inequality
\begin{equation}
\omega_{\max}\leq\operatorname*{Tr}X=\sum_{j=1}^{n}\omega_{j}\leq
n\lambda\omega_{\max}\label{maxnmax}%
\end{equation}
we have
\begin{equation}
\frac{1}{2}h\leq c(\Omega_{\mathrm{F}})\leq\frac{nh}{2}.\label{nh}%
\end{equation}

An immediate consequence of the inequality $\frac{1}{2}h\leq c(\Omega
_{\mathrm{F}})$ is that the Fermi ellipsoid $\Omega_{\mathrm{F}}$ of a
generalized coherent state always contains a quantum blob; this is of course
consistent with the uncertainty principle.

Notice that when all the eigenvalues $\omega_{j}$ are equal to a number
$\omega$ then $c(\Omega_{\mathrm{F}})=nh/2$; in particular when $n=1$ we have
$c(\Omega_{\mathrm{F}})=h/2$ which is exactly the action calculated along the
trajectory corresponding to the ground state. This observation leads us to the
following question: what is the precise geometric meaning of formula
(\ref{cwf})? Let us come back to the interpretation of the ellipsoid defined
by the inequality (\ref{ferx}). We have seen that the symplectic eigenvalues
of the matrix $%
\begin{pmatrix}
X & 0\\
0 & X
\end{pmatrix}
$ are precisely the eigenvalues $\omega_{j}$, $1\leq j\leq n$, of the
positive-definite matrix $X$. It follows that there exist linear symplectic
coordinates $(x^{\prime\prime},p^{\prime\prime})$ in which the equation of the
ellipsoid $\Omega_{\mathrm{F}}$ takes the normal form%
\begin{equation}
\sum_{j=1}^{n}\omega_{j}(x_{j}^{\prime\prime2}+p_{j}^{\prime\prime2})\leq
\sum_{j=1}^{n}\hbar\omega_{j}\label{omf1}%
\end{equation}
whose quantum-mechanical interpretation is clear: dividing both sides by two
we get the energy shell of the anisotropic harmonic oscillator in its ground
state. Consider now the planes $\mathcal{P}_{1},\mathcal{P}_{2},..,\mathcal{P}%
_{n}$ of conjugate coordinates $(x_{1},p_{1})$, $(x_{2},p_{2})$,...,
$(x_{n},p_{n})$. The intersection of the ellipsoid $\Omega_{\mathrm{F}}$ with
these planes are the circles
\begin{gather*}
C_{1}:\omega_{1}(x_{1}^{\prime\prime2}+p_{1}^{\prime\prime2})\leq\sum
_{j=1}^{n}\hbar\omega_{j}\\
C_{2}:\omega_{2}(x_{2}^{\prime\prime2}+p_{2}^{\prime\prime2})\leq\sum
_{j=1}^{n}\hbar\omega_{j}\\
\cdot\cdot\cdot\cdot\cdot\cdot\cdot\cdot\cdot\cdot\cdot\cdot\cdot\cdot
\cdot\cdot\cdot\\
C_{n}:\omega_{n}(x_{n}^{\prime\prime2}+p_{n}^{\prime\prime2})\leq\sum
_{j=1}^{n}\hbar\omega_{j}.
\end{gather*}
Formula (\ref{cwf}) says that $c(\Omega_{\mathrm{F}})$ is the area of the
circle $C_{j}$ with smallest radius, and this corresponds to the index $j$
such that $\omega_{j}=\omega_{\max}$. This is of course perfectly in
accordance with the definition of the Hofer--Zehnder capacity
$c^{_{\mathrm{HZ}}}(\Omega_{\mathrm{F}})$ since all symplectic capacities
agree on ellipsoids. We are now led to another question: is there any way to
describe topologically Fermi's ellipsoid in such a way that the areas of every
circle $C_{j}$ becomes apparent? The problem with the standard capacity of an
ellipsoid is that it only \textquotedblleft sees\textquotedblright\ the
smallest cut of that ellipsoid by a plane of conjugate coordinate. The way out
of this difficult lies in the use of the Ekeland--Hofer capacities
$c_{j}^{\mathrm{EH}}$ discussed above. To illustrate the idea, let us first
consider the case $n=2$; it is no restriction to assume $\omega_{1}\leq
\omega_{2}$. If $\omega_{1}=\omega_{2}$ then the ellipsoid%
\begin{equation}
\omega_{1}(x_{1}^{\prime\prime2}+p_{1}^{\prime\prime2})+\omega_{2}%
(x_{2}^{\prime\prime2}+p_{2}^{\prime\prime2})\leq\hbar\omega_{1}+\hbar
\omega_{2}\label{omf2}%
\end{equation}
is just the ball $B^{2}(\sqrt{2\hbar})$ whose symplectic capacity is
$2\pi\hbar=h$. Suppose now that $\omega_{1}<\omega_{2}$. Then the
Ekeland--Hofer capacities are the numbers%
\begin{equation}
\frac{\pi\hbar}{\omega_{2}}(\omega_{1}+\omega_{2}),\frac{\pi\hbar}{\omega_{1}%
}(\omega_{1}+\omega_{2}),\frac{2\pi\hbar}{\omega_{2}}(\omega_{1}+\omega
_{2}),\frac{2\pi\hbar}{\omega_{1}}(\omega_{1}+\omega_{2}),....\label{seq}%
\end{equation}
and hence
\[
c_{1}^{\mathrm{EH}}(\Omega_{\mathrm{F}})=c(\Omega_{\mathrm{F}})=\frac{\pi
\hbar}{\omega_{2}}(\omega_{1}+\omega_{2}).
\]
What about $c_{2}^{\mathrm{EH}}(\Omega_{\mathrm{F}})$? A first glance at the
sequence (\ref{seq}) suggests that we have
\[
c_{2}^{\mathrm{EH}}(\Omega_{\mathrm{F}})=\frac{\pi\hbar}{\omega_{1}}%
(\omega_{1}+\omega_{2})
\]
but this is only true if $\omega_{1}<\omega_{2}\leq2\omega_{1}$ because if
$2\omega_{1}<\omega_{2}$ then $(\omega_{1}+\omega_{2})/\omega_{2}<(\omega
_{1}+\omega_{2})/\omega_{1}$ so that in this case
\[
c_{2}^{\mathrm{EH}}(\Omega_{\mathrm{F}})=\frac{\pi\hbar}{\omega_{2}}%
(\omega_{1}+\omega_{2})=c_{1}^{\mathrm{EH}}(\Omega_{\mathrm{F}}).
\]
The Ekeland--Hofer capacities thus allow a geometrical classification of the eigenstates.

\section{Fermi Function and Excited States}

The generalized coherent states can be viewed as the ground states of a
generalized harmonic oscillator, with Hamiltonian function a homogeneous
quadratic polynomial in the position and momentum coordinates:%
\[
H(x,p)=\sum_{i,j}a_{ij}p_{i}p_{j}+b_{ij}p_{i}x_{j}+c_{ij}x_{i}x_{j}.
\]
Such a function can always be put in the form%
\begin{equation}
H(z)=\frac{1}{2}Mz\cdot z\label{quaham}%
\end{equation}
where $M$ is a symmetric matrix (the Hessian matrix, i.e. the matrix of second
derivatives, of $H$). We will assume for simplicity that $M$ is
positive-definite; we can then always bring it into the normal form
\[
K(z)=\sum_{j=1}^{n}\frac{\omega_{j}}{2}(x_{j}^{2}+p_{j}^{2})
\]
using a linear symplectic transformation of the coordinates (symplectic
diagonalization): there exists a symplectic matrix $S$ (depending on $M$) such
that%
\begin{equation}
S^{T}MS=D=%
\begin{pmatrix}
\Lambda & 0\\
0 & \Lambda
\end{pmatrix}
\label{sms}%
\end{equation}
where $\Omega$ is a diagonal matrix whose diagonal entries consist of the
symplectic spectrum $\omega_{1},...,\omega_{n}$ of $M$. Thus, we have
$K(z)=H(Sz)$, or, equivalently,%
\begin{equation}
H(z)=K(S^{-1}z)\label{hk}%
\end{equation}
The ground state of each one-dimensional quantum oscillator
\[
\widehat{K}_{j}=\frac{\omega_{j}}{2}\left(  x_{j}^{2}-\hbar^{2}\frac
{\partial^{2}}{\partial x_{j}}\right)
\]
is the solution of $\widehat{K}_{j}\Psi=\frac{1}{2}\hbar\omega_{j}\Psi$, it is
thus the one-dimensional fiducial coherent state $(\pi\hbar)^{-1/4}%
e^{-x^{2}/2\hbar}$. It follows that the ground $\Psi_{0}$ state of
$\widehat{K}=\sum_{j}\widehat{K}_{j}$ is the tensor product of $n$ such
states, that is $\Psi_{0}(x)=(\pi\hbar)^{-n/4}e^{-|x|^{2}/2\hbar}$, the
fiducial coherent state (\ref{fid}). Returning to the initial Hamiltonian $H$
we note that the corresponding Weyl quantisation $\widehat{H}$ satisfies, in
view of Eqn. (\ref{hk}) the symplectic covariance formula $\widehat{H}%
=\widehat{S}\widehat{K}\widehat{S}^{-1}$where $\widehat{S}$ is any of the
\textit{two} metaplectic operators corresponding to the symplectic matrix $S$
(see the Appendix). It follows that the ground state of $\widehat{H}$ is given
by the formula $\Psi=\widehat{S}\Psi_{0}$.

The case of the excited states is treated similarly. The solutions of the
one-dimensional eigenfunction problem $\widehat{K}_{j}\Psi=E\Psi$ are given by
the Hermite functions
\begin{equation}
\Psi_{N}(x)=e^{-x^{2}/2\hbar}H_{N}(x/\sqrt{\hbar})\label{hermitre2}%
\end{equation}
with corresponding eigenvalues $E_{N}=(N+\frac{1}{2})\hbar\omega_{j}$. It
follows that the solutions of the $n$-dimensional problem $\widehat{K}%
\Psi=E\Psi$ are the tensor products
\begin{equation}
\Psi_{(N)}=\Psi_{N_{1}}\otimes\Psi_{N_{2}}\otimes\cdot\cdot\cdot\otimes
\Psi_{N_{n}}\label{tensor}%
\end{equation}
where $(N)=(N_{1},N_{2},...,N_{n})$ is a sequence of non-negative integers,
and the corresponding energy level is%
\begin{equation}
E_{(N)}=\sum_{j=1}^{n}(N_{j}+\tfrac{1}{2})\hbar\omega_{j}.\label{en}%
\end{equation}

This allows us to give a geometric description of all eigenfunctions of the
generalized harmonic oscillator, corresponding to a quadratic Hamiltonian
(\ref{quaham}). We claim \ that:

\begin{quotation}
\emph{Let} $\Psi$ \emph{be an eigenfunction of the operator}
\begin{equation}
\widehat{H}=(x,-i\hbar\nabla_{x})M(x,-i\hbar\nabla_{x})^{T}. \label{hhat}%
\end{equation}
\emph{The symplectic capacity of the corresponding Fermi blob} $\Omega_{F}$
\emph{is}
\begin{equation}
c(\Omega_{F})=\sum_{j=1}^{n}(N_{j}+\tfrac{1}{2})h \label{cof}%
\end{equation}
\emph{where the numbers} $N_{1},N_{2},...,N_{n}$ \emph{are the non-negative
integers} \emph{corresponding to the state }(\ref{tensor}) \emph{of the
diagonalized operator} $\widehat{K}=\sum_{j=1}^{n}\widehat{K}_{j}$.
\end{quotation}

\section*{APPENDIX: The Metaplectic Group}

The symplectic group $\operatorname*{Sp}(2n,\mathbb{R})$ has a covering group
of order two, the metaplectic group $\operatorname*{Mp}(2n,\mathbb{R})$. That
group consists of unitary operators (the metaplectic operators) acting on
$L^{2}(\mathbb{R}^{n})$. There are several equivalent ways to describe the
metaplectic operators. For our purposes the most tractable is the following:
assume that $S\in\operatorname*{Sp}(2n,\mathbb{R})$ has the block-matrix form%
\begin{equation}
S=%
\begin{pmatrix}
A & B\\
C & D
\end{pmatrix}
\text{ \ with \ }\det B\neq0. \tag{A1}\label{free}%
\end{equation}
The condition $\det B\neq0$ is not very restrictive, because one shows (de
Gosson \cite{principi,Birk,Birkbis}, Littlejohn \cite{Littlejohn}) that every
$S\in\operatorname*{Sp}(2n,\mathbb{R})$ can be written (non uniquely) as the
product of two symplectic matrices of the type above; moreover the symplectic
matrices arising as Jacobian matrices of Hamiltonian flows determined by
physical Hamiltonians of the type \textquotedblleft kinetic energy plus
potential\textquotedblright\ are of this type for almost every time $t$. To
the matrix (\ref{free}) we associate the following quantities (de Gosson
\cite{principi,Birk}):

\begin{itemize}
\item A quadratic form
\begin{equation}
W(x,x^{\prime})=\frac{1}{2}DB^{-1}x\cdot x-B^{-1}x\cdot x^{\prime}+\frac{1}%
{2}B^{-1}Ax^{\prime}\cdot x^{\prime}; \tag{A2}\label{w}%
\end{equation}
the matrices $DB^{-1}$ and $B^{-1}A$ are symmetric because $S$ is symplectic;

\item The complex number $\Delta(W)=i^{m}\sqrt{|\det B^{-1}|}$ where $m$
(\textquotedblleft Maslov index\textquotedblright) is chosen in the following
way: $m=0$ or $2$ if $\det B^{-1}>0$ and $m=1$ or $3$ if $\det B^{-1}<0$.
\end{itemize}

The two metaplectic operators associated to $S$ are then given by%
\begin{equation}
\widehat{S}\Psi(x)=\left(  \tfrac{1}{2\pi i\hbar}\right)  ^{n/2}\Delta(W)\int
e^{\frac{i}{\hbar}W(x,x^{\prime})}\Psi(x^{\prime})d^{n}x^{\prime}%
.\tag{A3}\label{meta}%
\end{equation}
The fact that we have two possible choices for the Maslov index is directly
related the fact that $\operatorname*{Mp}(2n,\mathbb{R})$ is a two-fold
covering group of the symplectic group $\operatorname*{Sp}(2n,\mathbb{R})$
\cite{Wiley,principi,Birk,fo89}.

The main interest of the metaplectic group in quantization questions comes
from the two following (related) \textquotedblleft symplectic
covariance\textquotedblright\ properties:

\begin{itemize}
\item Let $\Psi$ be a square integrable function (or, more generally, a
tempered distribution), and $S$ a symplectic matrix. We have%
\begin{equation}
W\Psi(S^{-1}z)=W(\widehat{S}\Psi)(z) \tag{A4}\label{wigcov}%
\end{equation}
where $\widehat{S}$ is any of the two metaplectic operators corresponding to
$S$;

\item Let $\widehat{H}$ be the Weyl quantisation of the symbol (= observable)
$H$. Let $S$ be a symplectic matrix Then the quantisation of $K(z)=H(Sz)$ is
$\widehat{K}=\widehat{S}^{-1}\widehat{H}\widehat{S}$ where $\widehat{S}$ is
again defined as above.
\end{itemize}

\end{document}